
\input amstex.tex
\documentstyle{amsppt}
\nologo
\NoBlackBoxes
\topmatter 
\title
Current algebras and light-cone quantization in $3+1$ dimensions
\endtitle
\author
Jouko Mickelsson
\endauthor
\affil
Department of Mathematics, University of Jyv\"askyl\"a,
SF-40100 Jyv\"askyl\"a 10, Finland
\endaffil
\date January 30, 1992
\enddate
\endtopmatter
\document

\define\a{\alpha}

\define\g{\gamma}

\redefine\O{\Omega}
\redefine\l{\lambda}

\define\RM{\Bbb R}
\define\CM{\Bbb C}

\define\gm{\bold g}

\define\<#1,#2>{\langle #1,#2\rangle}
\define\TR{\text{tr}}
\define\dep(#1,#2){\text{det}_{#1}#2}
\define\norm(#1,#2){\parallel #1\parallel_{#2}}
\redefine\Bbb{\bold}

ABSTRACT A polarization of the Lie algebras $Map(C, G)$ of gauge
transformations on the light-cone $C\subset\RM^4$ is introduced, using
splitting of the initial data on $C$ for the wave operator to positive
and negative frequencies. This generalizes the usual polarization
of affine Kac-Moody algebras to positive and negative frequencies and paves
the way to a generalization of the highest weight theory to the $3+1$
dimensional setting.

\vskip 0.7in
\bf 1. Introduction\rm

\vskip 0.3in
Groups of maps $Map(M,G)$ where $M$ is a compact manifold and $G$ is a
compact group form a natural generalization of loop groups $LG=Map(S^1,G).$
The latter have a well understood representation theory. The success of this
theory rests mainly on the fact that the group $LG$ polarizes naturally to
subgroups corresponding to positive or negative Fourier modes on the
circle. This allows the concept of highest weight representation for a
central extension of $LG.$

In higher dimension it is not immediately clear what would replace the
Fourier decomposition. Of  course, if for example $M=S^1\times X$ for some
manifold $X$, one could define the subgroups $N_{\pm}$ of $Map(M,G)$ consisting
of positive (resp. negative) Fourier modes on the circle $S^1.$ The group
has a central extension defined by averaging the central extension in the
$S^1$ direction over the manifold $X$, with respect to some measure on $X.$
However, it seems both for mathematical and physical reasons that this is not
the right concept. First the physical reason. In two dimensional quantum
field theory the central extension of $LG$ arises because of the normal
ordering
regularization of bilinear expressions involving quantum fields at the same
point. Similarly, in higher dimensions one has to introduce certain
regularizati
ons
for the bilinears in order to get finite expressions for the physical currents.
The net effect of the various regularizations is that the naive commutation
relations get modified by the so-called Schwinger terms. The essential
difference as compared to the two dimensional situation is that extension
defined by Schwinger terms is not a central one but an extension by an abelian
ideal, [M1, M2].

A mathematical motivation for studying the abelian extensions is a kind of
universality. The groups $Map(M,G)$ can be embedded into certain infinite-
dimensional linear groups $GL_p$, modelled by Schatten ideals of degree $2p>
\text{dim}M.$
The embedding comes from physics: It is the gauge action of $Map(M,G)$
on chiral fermions. When dim$M=1$ the index $2p=2$, meaning that the group is
modelled by Hilbert-Schmidt operators. All the groups $GL_p$ have topologically
nontrivial abelian extensions generalizing the central extension $p=1,$ [MR].
The embedding $Map(M,G)\subset GL_p$ gives an abelian extension of the former
group.

The Hilbert space $H_F$ of square integrable chiral fermion fields on $M$
splits
into subspaces $H_{\pm}$ corresponding to positive (resp. negative) eigenvalues
of the Dirac operator. This corresponds to the Fourier splitting on the circle
or real line. An element $g\in GL_p$ can be written as
$$g=\left(\matrix a&b\\c&d\endmatrix\right),\tag1.1$$
where $a:H_+\to H_+, \,\, b:H_-\to H_+$ etc. In a highest weight representation
there is a cyclic vector which is annihilated by the subgroup $\gm_+$
consisting
of matrices $g$ with $c=0.$ Let $G_{\CM}$ be the complexification of $G.$
When $M=S^1$ the intersection of $\gm_+$ with the
subgroup $LG_{\CM}$ consists of loops with positive frequency and likewise the
intersection of $\gm_-$ with $LG_{\CM}$ corresponds to negative frequencies.
Thus the natural polarization on $LG$ is induced by the splitting $H=H_+\oplus
H_-$ of the "one-particle space". The polarization defines a complex structure
on the subgroup of based loops. The complex structure makes available the
holomorphic Borel-Bott-Weil induction method for constructing highest weight
representations, [PS].

The situation is completely different in higher dimensions. The problem is
that the intersection of $\gm_+$ or $\gm_-$ with the subgroup $Map(M,G_{\CM})
\subset GL_p$ is normally very small; it contains only the constant maps. Thus
the splitting of $H$ does not induce a polarization on $Map(M,G).$ This is the
case when the physical space $M$ has a positive definite Riemannian metric.

The purpose of the present paper is to show that if one chooses as the
initial data surface not a space-like surface of the type time $t=const.$
but the light cone $C=\{x=(x_0,\bold x)| x_0^2-x_1^2-x_2^2-x_3^2=0\}$ the
polarization
in the initial data space, given by a Dirac or wave operator, indeed induces
a polarization on $\Cal G=Map(C,G).$ We shall specialize to the physically most
interesting four dimensional setting, but the discussion has an obvious
generalization to higher dimensions. The subgroups $\Cal G_{\pm}$ corresponding
to positive or negative frequencies of the Dirac (or wave) operator furthermore
have the important property that the Schwinger terms defining the abelian
extension vanish along directions of $\Cal G_{\pm}$. This is in accordance with
the central term $c$ of a Kac-Moody algebra, $c(x,y)=0$ if both $x$ and $y$
have positive (resp. negative) frequency. The results of this
paper make it possible to extend (at least parts of) the theory of highest
weight representations of loop groups into higher dimensions in a physically
motivated manner.

\vskip 0.4in
\bf 2. Light-cone initial data for the wave equation and the polarization\rm

\vskip 0.3in
Since the cone $C$ is not a manifold (it has a singularity at the vertex
$x=0$) we have to specify what we mean by a smooth function on $C.$ We
call a function $f$ smooth if it extends to a smooth function in an open
neighborhood of $C$ in $\Bbb R^4.$ Let $G$ be a compact Lie group. The
space $\Cal G$ of smooth maps $C\to G$ is a group under pointwise
multiplication. Its Lie algebra is the space $C\gm$ of smooth maps $C\to\gm,$
where $\gm$ is the Lie algebra of $G.$ The commutators are defined pointwise.

Let $\gm_{\CM}$ be the complexification of the Lie algebra $\gm.$
We shall identify the space $C\gm_{\CM}$ as the space of smooth $\gm_{\CM}$
valued solutions of the wave equation
\define\0d{\partial_0}
\define\1d{\partial_1}
\define\2d{\partial_2}
\define\3d{\partial_3}

$$(\0d^2-\1d^2-\2d^2-\3d^2)\phi=0\tag2.1$$
A solution of (2.1) is uniquely determined by its initial data ${\phi(0,\bold
x)
,
\0d\phi(0,\bold x)}$ on the surface $x_0=0.$ We shall define a norm in the
initi
al
data space by
$$||(\phi,\0d\phi)||^2=\int_{x_0=0}<\overline{\phi},(-\1d^2-\2d^2-\3d^2)^{1/2}
\phi> d^3\bold x + \int_{x_0=0} <\overline{\0d\phi},\0d\phi>
d^3\bold x\tag2.2$$
where the positive square root is chosen for the Laplacian and $<\cdot,\cdot>$
is an invariant bilinear form on the Lie algebra $\gm.$  We denote by $H$
the completion of the space of smooth initial data with compact support. The
norm extends to an obvious inner product in $H.$

A pair of smooth solutions $\phi,\phi'$ of (2.1) (with appropriate vanishing
conditions at infinity) defines a conserved current $j_{\mu}=
<\overline{\phi},\partial_{\mu}\phi'>-<\overline{\partial_{\mu}\phi},\phi'>.$
The bar means complex conjugation. The dual of the one-form
$v\mapsto v^{\mu}j_{\mu}$ is a three-form and its integral over a space-like
surface is denoted by $Q(\phi,\phi').$ The 'charge' $Q$ does not depend on the
choice
of the space-like surface. The light-cone $C$ can be thought of as a limiting
case of a space-like surface and therefore the charge is
$$Q(\phi,\phi')=\int_{C} (<\overline{\phi}, D\phi'>-<D\overline{\phi},\phi'>)
\frac{d^3\bold x}{x_0}\tag2.3$$
where $D=x^{\mu}\partial_{\mu}$ is the derivative along light rays. Note that
the integration must be carried out over both the future light-cone $C_+$ and
the past light-cone $C_-.$

The charge $Q$ is continuous with respect to the norm in $H$ and therefore it
extends to the whole of $H.$  Furthermore, it is easy to see that $Q$ is
nondegenerate. (The solutions of the wave equations with zero frequency are
not contained in the Hilbert space $H.$)

Any vector in $H$ can be Fourier expanded as
$$\phi(x)=\frac{i}{2(2\pi)^{3/2}}\int_{C} e^{ip\cdot x} \hat{\phi}(p)
\frac{d^3\bold p}{p_0}  \tag2.4$$
where the integration is over the light-cone in the momentum space. The norm
is then given by
$$||\phi||^2=\frac{1}{2}\int_{C} |\hat{\phi}(p)|^2 (1+(p_1^2+p_2^2+
p_3^2)^{1/2}) \frac{d^3\bold p}{p_0}\tag2.5$$
and the charge is
$$Q(\phi,\phi')=\frac{1}{2}\int_{C} <\overline{\hat{\phi}}(p),
\hat{\phi'}(p)> \frac{d^3\bold p}{p_0}.\tag2.6$$
We shall use the following distribution in $\Bbb R^4:$
$$D(x)=\frac{i}{2(2\pi)^3}\int_{C} e^{ip\cdot x}\frac{d^3\bold
p}{p_0}.\tag2.7$$
All distributions will be defined in the space of smooth rapidly decreasing
functions at infinity. An explicit formula for $D(x)$ is
$$D(x)=-\frac{1}{2\pi}\epsilon(x_0) \delta(x^2).\tag2.8$$

\define\xd{\frac{d^3\bold x}{x_0}} 
We shall need also the distribution $\xi(p,q)=Q(e^{ip\cdot x},e^{iq\cdot x}).$
{}From the definition we get
$$\align \xi(p,q)&=\int_{C} e^{i(q-p)\cdot x} ix\cdot(p+q)
\frac{d^3\bold x}{x_0}\\
&=\frac{d}{d\l}\vert_{\l=1}\int e^{i(\l q-p)\cdot x}\xd
-\frac{d}{d\l}\vert_{\l=1}\int e^{i(q-\l p)\cdot x} \xd \\
&= 2i(2\pi)^2\frac{d}{d\l}_{\l=1}\left[ \epsilon(\l q_0-p_0)\delta((\l q-p)^2)
-2i(2\pi)^2\epsilon(q_0-\l p_0)\delta((q-\l p)^2)\right] \\
&=2i(2\pi)^2(q_0+p_0)\delta(q_0-p_0)\delta((q-p)^2)+
2i(2\pi)^2\epsilon(q_0-p_0) 2(q^2-p^2)
\delta'((p-q)^2).\tag2.9\endalign$$

Note in particular that $\xi(p,q)$ has support only on the set of light like
separated points $(p,q).$ Another important property of $\xi$ is that its
restriction to $C\times C$ is proportional to the Dirac $\delta$ distribution
on $C;$ this latter property can be read of also from the formula (2.6).

\proclaim{Lemma 1} Let $p,q$ be both on the future (resp. past) light-cone.
Then the distribution $\xi(k,p+q)$, when restricted to $k\in C$, has support
only on the future (resp. past) light-cone. \endproclaim

\demo{Proof} Suppose e.g. that $p,q\in C_+.$ Now $p+q$ is a future pointing
time
or light like vector and thus $k\cdot(p+q)$ is nonpositive for $k\in C_-.$
It follows that $(k-(p+q))^2\geq 0$ and the equality sign can occur only when
$p+q$ is light like, that is, when $p$ and $q$ are linearly dependent. When
$(k-(p+q))^2>0$ we know that $\xi(k,p+q)=0$ and in the latter case ($p+q$
on the light-cone) $\xi(k,p+q)$ is proportional to $\delta(k-(p+q))$ which is
zero when $0\neq k \in C_+.$
\enddemo
\define\Cg{{\gm_{\Bbb C}}}
\proclaim{Proposition 2} The subspaces $\bold b_{\pm}$ consisting of
restrictions
of solutions of the wave equation (2.1) to the light cone $C$, corresponding to
either frequencies $p_0\geq 0$ or $p_0\leq 0$, are subalgebras of $C\Cg_{\CM}$
under pointwise commutators. \endproclaim
\demo{Proof} The $k$:th Fourier component of a function $\phi:C\to \Cg_{\CM}$
(with $k\in C$) is given by the integral
$$ \hat{\phi}(k)=\frac{1}{(2\pi)^{3/2}}Q(e^{ik\cdot x}, \phi)\tag2.10$$
If a pair of solutions $\phi,\phi'$ has support only on the positive light-cone
in momentum space then their pointwise commutator is an integral over Fourier
modes $e^{ip\cdot x}$ where $p$ is a time like vector and $p_0\geq 0.$ From
the Lemma follows then that $[\phi,\phi']$, when restricted to the light cone,
contains only momenta $k\in C_+.$ \enddemo

In the case of a loop group the structure constants of the Lie algebra $L\gm_{
\CM}$ are simple. A basis is given by the Fourier modes $T_{a,n}=T_a
e^{in\phi},
$ $n\in \Bbb Z$ and the $T_a$'s form a basis of $\gm.$ One has then
$$[T_{a,n},T_{b,m}]= c_{ab}^f \l_{nm}^k T_{f,k}\tag2.11$$
where the $c$'s are the structure constants of $\gm$ and $\l_{nm}^k= \delta(n+
m-k).$ In the case of the Lie algebra $C\Cg$ the structure constants
$\l_{nm}^k$
are replaced by the distribution $\xi(p,q),$
$$[T_a e^{ip\cdot x},T_b e^{iq\cdot x}]=c_{ab}^f T_f \frac{i}{2(2\pi)^3}
\int_C \xi(k,p+q) e^{ik\cdot x} \frac{d^3\bold k}{k_0}\tag2.12$$
where both sides should be understood as distributions in $(p,q).$

We shall work out more explicitly the case of spherically symmetric functions.
A spherically symmetric solution of the wave equation is an integral over the
frequency $\l$ of the elementary complex valued solutions
$$\phi_{\l} (x)= e^{ix_0\l} \frac{\text{sin}(\l r)}{r}.\tag2.13$$
Here $r^2=x_1^2+x_2^2+x_3^2.$
The restriction of $\phi_{\l}$ to the light cone depends only on the time
coordinate $x_0$, $\phi_{\l}(x_0)=\phi_{\l}(x_0,r=|x_0|)=(e^{2i\l x_0}-1)/
2ix_0.$ We can write
$$\phi_{\l}\phi_{\l'}=\int \phi_{\mu} \a(\mu,\l,\l')d\mu\tag2.14$$
where, by simple Fourier analysis,
$$\a(\mu,\l,\l')= \epsilon(\l+\l'-\mu)-\epsilon(\l-\mu)-\epsilon(\l'-\mu)
+\epsilon(-\mu),\tag2.15$$
where $\epsilon(x)=+1$ for $x\geq 0$ and $\epsilon(x)=-1$ for $x<0.$ Note
that if $\l,\l'$ are both positive (resp. negative) then $\a$ is nonzero only
for positive (resp. negative) values of $\mu.$

In the general case a solution of the wave equation can be expanded in terms of
the functions
$$\phi_{\l,j,m}=e^{i\l x_0} X_j(\l r) Y_{jm}(\theta,\psi)\tag2.16$$
where the $Y_{jm}$'s are the normalized spherical harmonics ($j=0,1,2,\dots$
and
$m=-j,-j+1,\dots, j$) and the $X_j$'s are related to Bessel functions, [CH],
$$X_j(z)= J_{j+\frac12}(z)/\sqrt(z).\tag2.17$$
These solutions are smooth everywhere in $\Bbb R^4.$ Their restrictions to $C$
are smooth functions, evaluated by setting $r=|x_0|.$ In order to obtain
explicit commutation relations for the Lie algebra $C\gm$ one has to expand
$$\phi_{\l,j,m}\phi_{\l',j',m'}=\sum_{\ell,n}\int \phi_{\mu,\ell,n}
\a(\mu,\ell,n;\l,j,m;\l',j',m') d\mu.\tag2.18$$
In Fourier-Bessel analysis one can always expand a continuous function on the
real half line $[0,\infty[$, subject to the condition (2.21) below, in terms of
Bessel functions of a given order $\ell,$
$$f(x)=\int \l J_{\ell}(\l x)\hat{f}(\l) d\l,\tag2.19$$
where the coefficients are given by
$$\hat{f}(\l)=\int f(x) J_{\ell}(\l x)dx,\tag2.20$$
where we consider only functions satisfying
$$\int_{0}^{\infty} x|f(x)|^2 dx < \infty.\tag2.21$$
Since a smooth function on $S^2$ can also be expanded in terms of spherical
harmonics, we can easily show that a product of the functions $\phi_{\l,j,m}$
on the light cone can indeed be expanded as a sum and integral over the
solutions themselves.

We do not have an explicit formula for the distributional coefficients $\a,$
but
they can in principle be determined from the integral
$$\a(\mu,\ell,n;\l,j,m;\l',j',m')=  Q(\phi_{\mu,\ell,n},\,\, \phi_{\l,j,m}
\phi_{\l',j',m'}). \tag2.22$$
This follows from the orthogonality and completeness relations for the
spherical
harmonics and Bessel functions. The latter are, [W],
$$\int_{0}^{\infty} x J_{\ell} (\l x) J_{\ell}(\l' x) dx =\delta(\l-\l').
\tag2.23$$
By conservation of charge we have
$$\align Q(\phi_{\l,j,m},\,\phi_{\l',j',m'})&=\int_{x_0=0} (\l+\l')
\overline{\phi_{\l,j,m}}\phi_{\l',j',m'} r^2 dr d\Omega \\
&=\delta(j-j')\delta(m-m') \int_{0}^{\infty} rJ_j (\l r) J_j(\l' r)dr\\
&=\delta(j-j')\delta(m-m') \delta(\l-\l').  \tag2.24\endalign$$

\vskip 0.4in
\bf 3.  Action on chiral fermions \rm

\vskip 0.3in
We shall consider the space of solutions of the Dirac-Weyl equation
$$\sum_{0}^{3} \sigma^{\mu} \partial_{\mu} \psi =0,\tag3.1$$
where the $\sigma$'s are complex hermitian Pauli matrices, $\sigma_0=1,$
$\sigma_k^2=1,$ and $\sigma_1\sigma_2=-\sigma_2\sigma_1=\sigma_3,$ and cyclic
permutations. A smooth solution in Minkowski space is completely determined
by its Cauchy data on a space-like hyperplane. Alternatively, one can give
initial data on the light-cone $x^2=0.$ In the latter case there is a
constraint between the two components of the spinor field. This is understood
when looking at the Lorentz invariant inner product between solutions of
(3.1) which are square integrable on a space-like hypersurface $S.$ The inner
product is
$$(\psi,\psi')= \int_{S} \psi^*\sigma^{\mu}\psi' n_{\mu} d^3\bold x,\tag3.2$$
where $n$ is the future pointing unit normal to the surface $S.$ In the
limiting case when the surface $S$ approaches the light-cone one gets
$$(\psi,\psi')=\int_C \psi^* x^{\mu}\sigma_{\mu} \psi'\frac{d^3x}{x_0}.
\tag3.3$$
On the cone $C$ the matrix $P(x)=\frac{1}{2x_0}x_{\mu}\sigma^{\mu}$ is
degenerate, its eigenvalues are $0$ and $1.$ Thus only the one-component
field $P(x)\psi$ contributes to the norm of the solution $\psi$ and therefore
$\psi$ is completely determined by the values of $P\psi$ on $C,$ the second
component being given by the field equation.

The kernel of $P(x)-1$ in $\CM$ is a complex line depending only on the
direction of the vector $(x_1,x_2,x_3).$ In this way we obtain a complex
line bundle $E$ over $S^2.$ The independent initial data are sections of the
line bundle, with coefficients in the space of complex valued functions on
the real line $\RM,$ corresponding to the light ray parametrized by $\frac{
\bold x}{x_0}\in S^2.$  The line bundle $E$ is the basic complex line bundle
associated to the fibering $SU(2)\to SU(2)/S^1=S^2.$

Suppose from now on that $\psi$ takes values in $\CM^2\times \CM^N$ and a
compact gauge group acts on the latter components in the tensor product.
Because of the constraints on $C$, we cannot freely gauge transform an
initial data through the pointwise multiplication $\psi'(x)=g(x)\psi(x)$
where $g$ takes values in (a representation of)  $G.$ However, we can
multiply sections of the bundle $E\otimes Map(\RM, \CM^N)$ pointwise by
a gauge transformation.

Let $H_F$ be the Hilbert space of square-integrable sections in $E\otimes
Map(\RM,\CM^N).$ This space splits to subspaces $H_{\pm},$ defined by the
nonnegative (resp., negative) eigenvalues of the Hamiltonian $h=i\sum \sigma^k
\partial_k$ on the space of solutions of (3.1) (identified as the space of
initial data $H_F$).

\proclaim{Proposition 3} A gauge transformation in $ \Cal G_+$ maps $H_+$
onto itself and similarly an element of $\Cal G_-$ maps $H_-$ onto itself.
\endproclaim
\demo{Proof} The initial data on $C$ for a solution corresponding to energy
$\l$, angular momentum $j=\ell+1/2,$ and the third component of angular
momentum equal to $j_3=m+1/2$ ($\ell=0,1,2,\dots$ and $m=-\ell,-\ell+1,
\dots,\ell$) are functions of time $t=x_0$ and of spherical angles $\theta,
\phi$ parametrizing the point $\bold x/x_0.$ An explicit formula is
$$\psi_{\l,j,j_3}=f_{\l,\ell+1/2}(t)\left(\matrix Y_{\ell,m}\\ \g_+ Y_{\ell,
m+1}\endmatrix\right)+f_{\l,\ell+3/2}(t)\left(\matrix Y_{\ell+1,m}\\ \g_-
Y_{\ell+1,m+1}\endmatrix\right),\tag3.4$$
where $\g_+=\sqrt{\frac{\ell-m}{\ell+m+1}},\, \g_-=-\sqrt{\frac{\ell+m+2}
{\ell-m+1}},$ $Y_{\ell,m}$'s are spherical harmonics, and
$$f_{\l,p}(t)=\frac{e^{i\l t}}{\sqrt{r}}j_{p}(\l r),\tag3.5$$
where $r=|t|$ and the $j_{\l,p}$'s are, up to a normalization constant,
again Bessel functions of order $p.$

The projection of $\psi$ onto the complex line $P(x)\CM$, as a function of
$t$, is a linear combination of the functions $f_{\l,\ell+1/2}$ and $f_{\l,
\ell+3/2}.$ Both of these functions have only positive Fourier components
when $\l$ is positive; this follows immediately from the integral
representation
$$J_n(z)= \frac{z^n}{2^{n+1} n!}\int_0^{\pi} \text{cos}(z \text{cos}\theta)
\text{sin}^{2n+1} \theta d\theta\tag3.6$$
for Bessel functions. Thus the product of $\psi$ with a positive energy
wave function $\phi$  produces a function with only nonnegative Fourier
componen
ts.
On the other hand, a section corresponding to negative energy contains only
negative Fourier components (as a function of time on $C$) and therefore
$\phi\psi$ cannot contain negative energy components. \enddemo

\proclaim{Proposition 4} Let $\phi:C\to G$ be a smooth bounded function such
that $p_0^{n}\hat\phi(p)\to 0$ for $p_0\to\pm\infty,$ where $\hat\phi$ is the
Fourier transform of $\phi.$ Write
the operator corresponding to a pointwise multiplication by $\phi$ in $H_F$ as
$$T(\phi)=\left(\matrix a&b\\c&d\endmatrix\right).$$
Then the off-diagonal blocks are in the Schatten ideal $L_4$, that is,
$\TR |c|^4<\infty,\TR |b|^4<\infty.$ \endproclaim
\demo{Proof} Here it is more convenient to work in the momentum basis.
The solution of the Weyl equation with 4-momentum $p\in C$ is
$$\psi_p(x)= \left(\matrix \frac{p_1+ip_2}{\sqrt{2p_0(p_0-p_3)}}\\
\sqrt{\frac{p_0-p_3}{2p_0}}\endmatrix\right) e^{ip\cdot x}.\tag3.7$$
The norm of a square-integrable solution
$$\psi=\frac{1}{(2\pi)^{3/2}}\int_C \psi_p\a(p)\frac{d^3\bold p}{p_0}
\tag3.8$$
is given by the Lorentz invariant integral
$$||\psi||^2=\int_C \a(p)\sigma^{\mu}\a(p)p_{\mu}\frac{d^3\bold
p}{p_0}.\tag3.9$$
We need an estimate for the integral
$$I(\phi)=\int_X |T(p,p')|^4 d^3\bold p d^3\bold p',$$
where $X=C_-\times C_+$ or $X=C_+\times C_-$ and $T(p,q)$ is the matrix element
of a multiplication operator in the momentum basis. To start with, let $\phi(x)
=e^{iq\cdot x}$ with $q\in C_+.$ The integral $I$ is finite if and only if the
integral
$$I'(\phi)=\int_{C_-}||\pi_{+}\phi\psi_p||^4 d^3\bold p$$
is finite, where $\pi_{\pm}$ are projectors on the subspaces $H_{\pm}.$

A section corresponding to positive (negative) energy is characterized by the
property that as a function of time $t=x_0$ it contains only positive
(negative)
Fourier components. Thus we can write
$$\pi_+ e^{iq\cdot x}\psi_p= \cases e^{iq\cdot x}\psi_p \text{ when $(p+q)\cdot
x/t \geq 0$}\\  0 \text{ when $(p+q)\cdot x/t <0$}\endcases.$$
Denote by $A(p,q)$ the area on $S^2$ of the set of points $\bold x/t$ with
$(p+q)\cdot x/t \geq 0.$ By separating the angular and time variables in the
definition of the norm in $H_F$ we observe that for large $p_0$ the integral
$I'$ behaves like
$$I''(\phi)=\int A(p,q)^4 d^3\bold p.$$

Denote $\hat x=x/x_0.$
An estimate for the area $A$ can be reduced from the inequalities
$$(p+q)\cdot \hat x\geq 0 \text{ or } 1-\hat{\bold p}\cdot\hat{\bold x}\leq
q\cdot \hat x/p_0.$$
Denoting by $\theta$ the angle between $\hat{\bold x}$ and $\hat{\bold p}$ we
get
$$\text{cos}\theta \geq 1-|\frac{q_0}{p_0}|.$$
For large $p_0$ the area $A(p,q)$ behaves like $\theta^2 \sim |\frac{q_0}{
p_0}|$ and therefore the integrand in $I$'' can be substituted by
$q_0^4/p_0^4.$
The integral of this with the measure $d^3\bold p=p_0^2 dp_0 d\O$ converges
at $p_0\to\infty.$

The general case is handled using a Fourier decomposition of $\phi$ in terms
of the functions $e^{iq\cdot x}$ on the light-cone; by the assumption, the
integral of $\hat\phi$ with $q_0^4$ is converging.  \enddemo

In quantum theory the gauge transformations on $C$ are naivly effected by
the component $x^{\mu}j_{\mu}=:\psi(x)x^{\mu}\sigma_{\mu}\psi(x):$ of the
current operator $j_{\mu}.$ Here the dots refer to a normal ordering with
respect to the free fermionic Hamiltonian. However, in $3+1$ dimensions even
the normal ordered operators are ill-defined in the fermionic Fock space.
In general, applying the smeared current $\int j_{\mu}(x) f(x) dx$ to a
vector in the Fock space produces a vector of infinite norm. This is an
indication of the fact that a gauge transformation tends to send a vector in
a representation of canonical anticommutation relations (CAR) to a vector in
a different, \it inequivalent\rm, representation of CAR.

One can make sense of the gauge transformations as unitary maps between
different CAR representations, [M2]. Alternatively, one can consider the
gauge transformations as sesquilinear forms in a Fock space, [R], with a
suitably defined twisted product, [L].

The Proposition 4 shows that the gauge transformations on the light cone
can be embedded to the group $GL_2$ consisting of operators $g=\left(
\smallmatrix a&b\\c&d\endsmallmatrix \right)$ in $H_F=H_+\oplus H_-$ with
$b,c\in L_4.$ The group $GL_2$ acts through an abelian extension $\widehat{
GL}_2$ in a bundle $\Cal F$ of fermionic Fock spaces $\Cal F_W$ parametrized
by certain representations of CAR. The base space is an infinite-dimensional
Grassmannian $Gr_2$ consisting of closed subspaces $W\subset H_F$ such that the
orthogonal projection $W\to H_+$ is Fredholm and the projection $W\to H_-$ is
in $L_4.$ To each $W\in Gr_2$ there corresponds an irreducible representation
of CAR characterized by the existence of a vacuum vector $|W>$ with the
property
$$a^*(u)|W>=0=a(v)|W> \forall u\in H_-, v\in H_+\tag3.10$$
where the CAR is generated by creation operators $a^*(u)$ and annihilation
operators $a(v)$ with the only nonvanishing anticommutation relations
$$a^*(u)a(u')+a(u')a^*(u)=<u,u'>  \text{ with } u,u'\in H.\tag3.11$$
Two representations parametrized by $W,W'$ are equivalent if and only if
the projection of $W'$ to $W^{\perp}$ is Hilbert-Schmidt, [A].

The group $GL_2$ acts in the base in a natural way but in the case of Weyl
fermions one has to "twist" the action in the total space $\Cal F,$ [M2].
Infinitesimally, this means that the normal operator commutators are replaced
by the commutators, [MR],
$$[X,Y]_c = [X,Y]+ c_2(X,Y;W),\tag3.12$$
where the 2-cocycle $c_2$ is a function of the "background" W,
$$c_2(X,Y;W)=\frac18\TR (\epsilon-F) [[\epsilon,X],[\epsilon,Y]],\tag3.13$$
where $F:H\to H$ is the linear operator characterized by $Fu=u$ for $u\in W$
and $Fu=-u$ for $u\in W^{\perp},$ and $\epsilon$ is the sign of the free
Hamiltonian $h,$  $\epsilon=h/|h|$ (we use the convention that the zero
eigenvalue of $h$ corresponds to sign $+1$).

The cocycle $c_2$ has the property that it vanishes if both $X$ and $Y$
belong to the subalgebra $\gm_+\subset {\bold gl}_2$ (resp., the
subalgebra $\gm_-$) characterized by vanishing of the off-diagonal block
$c$ (resp., $b$). Now $\bold b_+\subset \gm_+$ and $\bold b_-\subset \gm_-,$
threfore we have:
\proclaim{Proposition 5} The restriction of the cocycle $c_2$ to either of the
subalgebras $\bold b_{\pm}\subset C\gm_{\CM}$ vanishes. \endproclaim

Note that the restriction of $c_2$ to $C\gm_{\CM}$ differs from the simple
\it local \rm cocycle
$$c_2'(X,Y;f)=\int_C \TR (df\,f^{-1})^2(XdY-YdX),\tag3.14$$
where the trace is computed in a finite-dimensional representation of $G$ and
$f\in \Cal G,$ [M1, Chapter 4].

Although we have an unitary action of $\widehat{GL}_2$ between fibers, we
do not have an unitary representation of the group in a (separable) Hilbert
space since there is no invariant measure on the base, [Pi].

In order to define the action of the other three components of the 4-current
$j_{\mu}$ we have to find an embedding to the the Lie algebra of $GL_2.$
Let $n\in \RM^4$ be any time-like unit vector and let $S(n)$ be the plane in
$\RM^4$ with unit normal equal to $n.$ For a smooth function $f$ on $S(n)$
define formally
$$j(n,f)=\int :\psi(x)n^{\mu}\sigma_{\mu}\psi(x): f(x) dS(n),\tag3.14$$
where $S(n)$ is the translation invariant measure induced by the measure
$d^4x$ on $\RM^4.$ We would like to interprete the $j(n,f)$'s as generators of
gauge transformations for initial data on the \it space-like surface \rm
$S(n).$ In the non second quantized formalism there are no problems with
this. Again, we split the first quantized Hilbert space $H_F$ into $H_+\oplus
H_-$ with respect to the free Hamiltonian $h$, but now we are going to identify
the space of solutions of the Weyl equation as the space of square-integrable
initial data on $S(n),$ with the inner product
$$<\psi,\psi'>=\int \psi(x)^*n^{\mu}\sigma_{\mu}\psi(x) d^3\bold x.\tag3.15$$

Each smooth $G$ valued function $\phi$ on $S(n)$, with bounded derivatives,
defines an unitary operator
in $H$ by pointwise multiplication. This operator $T(\phi)$ belongs to
$GL_2.$ The proof is even simpler than in the case of the cone. One has
to show that the commutator $[\epsilon, T(\phi)]\in L_4.$ This follows
from the fact that 1) the operator $[h,T(\phi)]$ is bounded (as a
multiplication
operator), and 2) the operator $|h|^{-1}$ raised to power 4 is trace class
(remember that the number of states in the volume $d^3\bold p$ is momentum
space is $p_0^2 d^3\bold p$).

The important point in the discussion above is that the current $n^{\mu}j_{\mu}
$ should be smeared with a smooth function on \it its characteristic surface
\rm
$S(n).$ For example, the space components $j_k$ integrated over the time slice
$x_0= const.$ are \it not \rm in the Lie algebra of $GL_2.$ By a direct
computation one can show that they do not even have compact off-diagonal
blocks (for a generic smearing function)!

\vskip 0.3in
\bf Acknowledments \rm The author has profited from fruitful discussions on
light-cone quantization with Sarada Rajeev, Roman Jackiw, and Vesa Ruuska.

\vskip 0.7in

\bf References \rm

\vskip 0.3in
[A] Araki, H.: Bogoliubov automorphisms and Fock representations of
canonical anticommutation relations. in: Contemporary Mathematics,
American Mathematical Society vol. 62 (1987)

[CH] Courant, R., D. Hilbert: \it Methods of Mathematical Physics. \rm
Interscience Inc., New York (1953)

[L] Langmann, E.: Fermion and boson current algebras in (3+1)-dimensions.
Proc. of the symposium "Topological and Geometrical Methods in Field
Theory", Turku, 1991. Eds. J. Mickelsson and O. Pekonen. World Scientific,
Singapore (1992)

[M1] Mickelsson, J.: \it Current Algebras and Groups. \rm Plenum Press,
New York and London (1989)

[M2] Mickelsson, J.: Commutator anomaly and the Fock bundle. Commun. Math.
Phys. \bf 127, \rm 285 (1990). On the Hamiltonian approach to commutator
anomalies in (3+1) dimensions. Phys. Lett. \bf B 241, \rm 70 (1990)

[MR] Mickelsson, J. and S. Rajeev: Current algebras in $d+1$ dimensions and
determinant bundles over infinite-dimensional Grassmannians. Commun. Math.
Phys. \bf 116, \rm 365 (1988)

[Pi] Pickrell, D.: On the Mickelsson-Faddeev extension and unitary
representations. Commun. Math. Phys. \bf 123, \rm 617 (1989)

[PS] Pressley, A. and G. Segal: \it Loop Groups. \rm Clarendon Press,
Oxford, 1986.

[R] Ruijsenaars, S.N.M.: Index formulas for generalized Wiener-Hopf operators
and Boson-Fermions correspondence in $2N$ dimensions. Commun. Math. Phys.
\bf 124, \rm 553 (1989)

[W] Watson, G.N: \it Theory of Bessel functions. \rm Cambridge University
Press, Cambridge, UK (1952)

\enddocument